\documentclass[aps,prx,preprint,superscriptaddress]{revtex4-1}

\usepackage[version=3]{mhchem} 
\usepackage[T1]{fontenc}       
\usepackage[version=3]{mhchem} 
\usepackage[T1]{fontenc}       
\usepackage{graphicx}          
\usepackage{dcolumn}           
\usepackage{bm}                
\usepackage{xcolor}            
\usepackage{amsmath}           
\usepackage{hyperref}          
\usepackage{tabularx}
\usepackage[english]{babel}

\begin{document}


\title{A new noncollinear ferromagnetic Weyl semimetal with anisotropic anomalous Hall effect}


\author{Hung-Yu~Yang}
\affiliation{Department of Physics, Boston College, Chestnut Hill, MA 02467, USA}
\author{Bahadur~Singh}
\affiliation{Department of Condensed Matter Physics and Materials Science, Tata Institute of Fundamental Research, Colaba, Mumbai 400005, India}
\affiliation{Department of Physics, Northeastern University, Boston, MA 02115, USA}
\author{Jonathan~Gaudet}
\affiliation{Institute for Quantum Matter and Department of Physics and Astronomy, The Johns Hopkins University, Baltimore, Maryland 21218, USA}
\author{Baozhu~Lu}
\affiliation{Department of Physics and Temple Materials Institute, Temple University, Philadelphia, PA 19122, USA}
\author{Cheng-Yi~Huang}
\affiliation{Department of Physics, Northeastern University, Boston, MA 02115, USA}
\author{Wei-Chi~Chiu}
\affiliation{Department of Physics, Northeastern University, Boston, MA 02115, USA}
\author{Shin-Ming~Huang}
\affiliation{Department of Physics, National Sun Yat-sen University, Kaohsiung 80424, Taiwan}
\author{Baokai~Wang}
\affiliation{Department of Physics, Northeastern University, Boston, MA 02115, USA}
\author{Faranak~Bahrami}
\affiliation{Department of Physics, Boston College, Chestnut Hill, MA 02467, USA}
\author{Bochao~Xu}
\affiliation{Department of Physics, University of Connecticut, Storrs, CT USA, 06269}
\author{Jacob~Franklin}
\affiliation{Department of Physics, University of Connecticut, Storrs, CT USA, 06269}
\author{Ilya~Sochnikov}
\affiliation{Department of Physics, University of Connecticut, Storrs, CT USA, 06269}
\affiliation{Institute of Material Science, University of Connecticut, Storrs, CT USA, 06269}
\author{David~E.~Graf}
\affiliation{National High Magnetic Field Laboratory, Tallahassee, FL 32310, USA}
\author{Guangyong~Xu}
\affiliation{NIST Center for Neutron Research, National Institute of Standards and Technology, Gaithersburg, MD 20899-6102, USA}
\author{Yang~Zhao}
\affiliation{NIST Center for Neutron Research, National Institute of Standards and Technology, Gaithersburg, MD 20899-6102, USA}
\author{Christina~M.~Hoffman}
\affiliation{Neutron Scattering Division, Oak Ridge National Laboratory, Oak Ridge, Tennessee 37831, USA}
\author{Hsin~Lin}
\affiliation{Institute of Physics, Academia Sinica, Taipei 11529, Taiwan}
\author{Darius~H.~Torchinsky}
\affiliation{Department of Physics and Temple Materials Institute, Temple University, Philadelphia, PA 19122, USA}
\author{Collin~L.~Broholm}
\affiliation{Institute for Quantum Matter and Department of Physics and Astronomy, The Johns Hopkins University, Baltimore, Maryland 21218, USA}
\author{Arun~Bansil}
\email{ar.bansil@northeastern.edu}
\affiliation{Department of Physics, Northeastern University, Boston, MA 02115, USA}
\author{Fazel~Tafti}
\email{fazel.tafti@bc.edu}
\affiliation{Department of Physics, Boston College, Chestnut Hill, MA 02467, USA}



\date{\today}

\begin{abstract}
An emerging frontier in condensed matter physics involves novel electromagnetic responses, such as the anomalous Hall effect (AHE), in ferromagnetic Weyl semimetals (FM-WSMs).
Candidate FM-WSMs have been limited to materials that preserve inversion symmetry and generate Weyl crossings by breaking the time-reversal symmetry.
These materials share three common features: a centrosymmetric lattice, a collinear FM ordering, and a large AHE observed when the field is parallel to the magnetic easy axis.
Here, we present CeAlSi as a new type of FM-WSM in which the Weyl nodes are stabilized by breaking the inversion symmetry, but their positions are tuned by breaking the time-reversal symmetry.
Unlike the other FM-WSMs, CeAlSi has a noncentrosymmetric lattice, a noncollinear FM ordering, and a novel AHE that is anisotropic between the easy and hard magnetic axes.
It also exhibits large FM domains that are promising for exploring both device applications and the interplay between the Weyl nodes and FM domain walls.
\end{abstract}


\maketitle

\section{Introduction}
Weyl nodes are protected linear crossings of two non-degenerate bands that lead to chiral relativistic quasiparticles~\cite{armitage_weyl_2018,bansil_colloquium_2016}.
In Weyl semimetals (WSMs), the presence of Weyl nodes at the Fermi level enables Berry phase engineering in the bulk, creates Fermi arcs on the surface, and leads to a host of emergent electromagnetic responses such as the topological Hall effect (THE) and the anomalous Hall effect (AHE)~\cite{tokura2017emergent,PhysRevLett.102.186602,PhysRevLett.106.156603,matsuno_interface-driven_2016,nagaosa_anomalous_2010,
yao_first_2004,yang_quantum_2011,burkov2014anomalous,ueda2018spontaneous,Nakatsuji_2015,nayak_large_2016,li_chiral_2019,destraz_magnetism_2020,yang_transition_2020}.
There are two main pathways for generating Weyl semimetals: breaking the inversion symmetry~\cite{weng_weyl_2015}, or the time-reversal symmetry~\cite{wan_topological_2011}.
The former approach yielded the original discovery of non-magnetic Weyl semimetals in TaAs family~\cite{huang2015weyl,lv_experimental_2015,xu_discovery_2015}.
The latter approach has recently led to the discovery of ferromagnetic Weyl semimetals (FM-WSMs) such as Co$_3$Sn$_2$S$_2$, Fe$_3$GeTe$_2$, and Co$_2$MnGa~\cite{wang_large_2018,liu_giant_2018,kim_large_2018,belopolski_discovery_2019}.
These FM-WSMs crystallize in a centrosymmetric lattice and exhibit collinear FM ordering.
They have been intensely studied due to a giant AHE that results from the Berry curvature around Weyl nodes, as confirmed by first-principle calculations~\cite{yao_first_2004,liu_giant_2018,
kim_large_2018,belopolski_discovery_2019}.

In this article, we introduce CeAlSi as a new type of FM-WSM that combines both routes mentioned above to generate Weyl nodes.
CeAlSi crystallizes in the noncentrosymmetric space group $I4_1md$, a point we confirm via our second-harmonic-generation (SHG) experiments and first-principles calculations.
The local $f$-moments of Ce$^{3+}$ are found to interact within the noncentrosymmetric lattice and lead to a noncollinear FM order.
The breaking of time-reversal symmetry in CeAlSi shifts the nodal positions and controls the magnitude of the AHE.
We observe two different AHE responses in this material by orienting the magnetic field along the easy and hard magnetic axes.
The lack of inversion symmetry, the in-plane noncollinear FM order, and the novel anisotropic AHE make CeAlSi a new FM-WSM candidate that is distinct from other FM-WSMs.

\section{\label{sec:results}Main Results}
\begin{figure*}
\centering
\includegraphics[width=\textwidth]{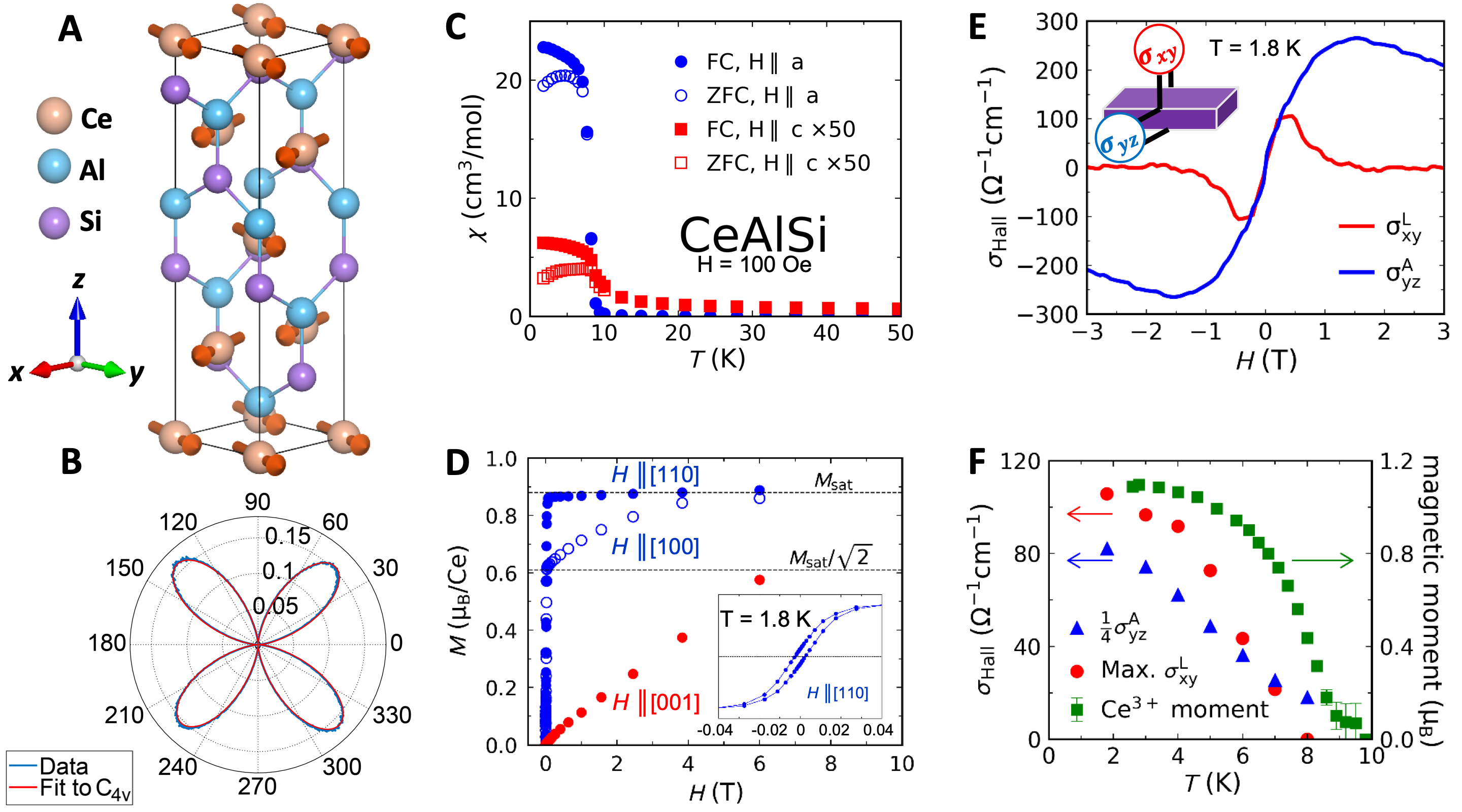}
\caption{\label{fig:UC}
\textbf{Non-collinear order, non-centrosymmetric structure, and anomalous Hall effect in CeAlSi.}
(\textbf{A}) Non-collinear FM order in the tetragonal unit cell of CeAlSi.
(\textbf{B}) Second-harmonic generation data refined in the noncentrosymmetric $C_{4\textrm{v}}$ point group.
(\textbf{C}) Anisotropic magnetic susceptibility as a function of temperature showing the in-plane easy-axis.
(\textbf{D}) Magnetization curves with the field oriented along the [110], [100], and [001] directions. Inset shows hysteresis due to FM domains with a coercive field of 70~Oe.
(\textbf{E}) Two distinct Hall responses are observed when a magnetic field is applied along the easy axis ($\sigma^A_{yz}$: anomalous Hall effect) or the hard axis ($\sigma^L_{xy}$: loop Hall effect).
(\textbf{F}) Evolution of the magnetic ordering parameter (the Ce$^{3+}$ moment), $\sigma^A_{yz}$, and $\sigma^L_{xy}$ with temperature.
}
\end{figure*}
Figure~\ref{fig:UC} summarizes our main results related to the discovery of a new noncentrosymmetric FM-WSM with an anisotropic AHE.
The body-centered tetragonal unit cell of CeAlSi (Fig.~\ref{fig:UC}A) contains two vertical mirror planes ($\sigma_\textrm{v}$) but lacks a horizontal mirror plane ($\sigma_\textrm{h}$), thus breaking the inversion symmetry.
The viability of an FM-WSM in such a structure (space group $I4_1md$) was first proposed by DFT calculations in CeAlGe~\cite{chang_magnetic_2018,xu_discovery_2017}; however, experiments reported an antiferromagnetic (AF) order instead of an FM order~\cite{hodovanets_single-crystal_2018,puphal_topological_2020,suzuki_singular_2019}.
On the contrary, our neutron diffraction and magnetization measurements show that CeAlSi hosts an FM order with net magnetization along the crystallographic [110] direction and an in-plane non-collinear spin texture as illustrated in Fig.~\ref{fig:UC}A.
Although the non-collinear FM order distinguishes CeAlSi from other FM-WSMs, we will show that the solid angle between the non-collinear spins does not change with magnetic field.
Thus, the AHE observed in CeAlSi is distinct from the THE in non-collinear magnets such as the Mn$_3$Sn and MnSi families~\cite{Nakatsuji_2015,nayak_large_2016,li_chiral_2019,PhysRevLett.102.186602,PhysRevLett.106.156603}.

An important structural detail is the possibility of site mixing between Al and Si, which could invalidate the proposal of CeAlSi being a noncentrosymmetric FM-WSM. 
Intersite mixing can restore the $\sigma_h$ mirror plane and change the space and point groups from noncentrosymmetic $I4_1md$ ($C_{4\textrm{v}}$) to centrosymmetric $I4_1/amd$ ($C_{4\textrm{h}}$).
Neither X-ray nor neutron diffraction can reliably distinguish between the two space groups, see Sec. M1 in the Supplemental Material for details.
However, SHG can discriminate between these two structures because the SHG signal predominantly originates from a bulk electric dipole in a noncentrosymmetric unit cell.
Figure~\ref{fig:UC}B shows a strong SHG signal ($\chi_{xxz}=200$~pm/V) that is commensurate with the pronounced signal in GaAs~\cite{bergfeld_second-harmonic_2003} and fits the point group $C_{4\textrm{v}}$.
Thus, we confirm the noncentrosymmetric space group $I4_1md$ as the correct structure, see Sec. M2 in the Supplemental Material for details.

CeAlSi is ferromagnetic with a strong magnetic anisotropy with an in-plane easy axis. As seen in Fig.~\ref{fig:UC}C, the in-plane magnetic susceptibility (blue) is 200 times larger than the out-of-plane susceptibility (red).
The field dependence of magnetization (Fig.~\ref{fig:UC}D) indicates that the [110] crystallographic direction as the easy axis.
A gradual saturation of the $M(H\|[100])$ curve from $M_{sat}(H\|[110])/\sqrt{2}$ to $M_{sat}(H\|[110])$ implies the presence of zero-field magnetic domains with $\mathbf{M}\|$~[110], [1-10], [-110], and [-1-10] directions.
%

Due to the in-plane easy-axis orientation, we expect to observe an AHE when the magnetic field is oriented in the $ab$-plane.
Figure~\ref{fig:UC}E confirms such an anomalous Hall conductivity (the step in $\sigma_{yz}^A$), but it also reveals an unexpected signal ($\sigma_{xy}^L$) which is observed when the field lies along the hard axis.
The superscript $L$ in $\sigma_{xy}^L$ stands for its loop-shape behavior.
Figure~\ref{fig:UC}F shows the parallel temperature dependence of $\sigma_{yz}^A$, $\sigma_{xy}^L$, and the magnetic order parameter determined by neutron diffraction, suggesting that both Hall responses are controled by the FM order.
We will examine these findings in detail in the remainder of this paper.

\section{\label{sec:neutron}In-Plane Noncollinear FM Order and Large FM Domains}
\begin{figure*}
\centering
\includegraphics[width=\textwidth]{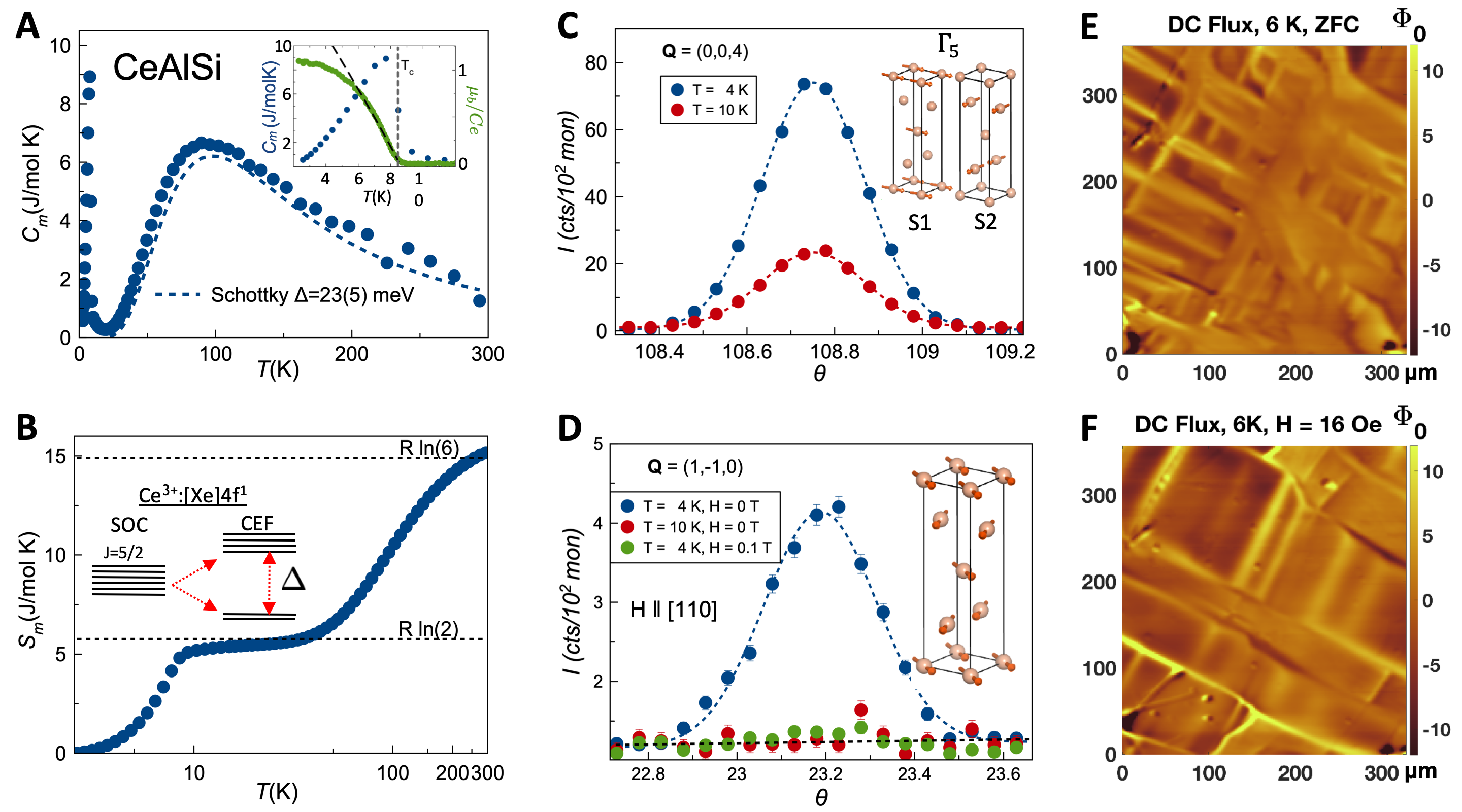}
\caption{\label{fig:NM}
\textbf{Magnetic structure.}
(\textbf{A}) Magnetic specific heat as a function of temperature with a fit to the crystal electric field (CEF) levels. Inset shows a magnified view of the FM transition and a fit to the temperature dependence of the order parameter (magnetic moment per Ce atom).
(\textbf{B}) Magnetic entropy as a function of temperature. Inset shows the CEF levels with a doublet ground-state.
(\textbf{C}) Neutron scattering Bragg peak at $\mathbf{Q}$~=~(004).
(\textbf{D}) Neutron scattering Bragg peak at $\mathbf{Q}$~=~(1-10) is observed below $T_C$~=~8.2(3)~K in zero-field and suppressed by applying a field of 0.1~T along the [110] direction.
(\textbf{E}) Scanning SQUID image of FM domains obtained at $T$~=~6~K under zero field.
(\textbf{F}) Large in-plane domains develop under a small in-plane field (16~Oe).
}
\end{figure*}
The magnetic heat capacity ($C_m$) of CeAlSi in Fig.~\ref{fig:NM}A exhibits a sharp FM transition at $T_C$~=~8.2(3)~K and a broad (Schottky) peak at 80~K due to the crystal electric field (CEF) splitting of Ce$^{3+}$ atomic levels.
As shown in Fig.~\ref{fig:NM}B, the $J=5/2$ sextet of Ce$^{3+}$ splits into a doublet ground-state and a quadruplet excited-state, leading to two plateaus at $R\ln(2)$ and $R\ln(6)$ in the magnetic entropy $S_m$.
From a fit to the $C_m$ data in Fig.~\ref{fig:NM}A, we estimate a gap of $\Delta$~=~25~meV between the doublet and the quadruplet, and identify the ground-state of CeAlSi as a Kramers doublet with effective spin-1/2.

The in-plane noncollinear FM order of CeAlSi was determined by neutron diffraction.
Figure~\ref{fig:NM}C shows the $\mathbf{Q}$~=~(004) peak corresponding to the FM ordering vector $\mathbf{k}$~=~(000).
The magnetic moment per Ce$^{3+}$ (order parameter) is extracted from the intensity of this peak and plotted as a function of temperature in the inset of Fig.~\ref{fig:NM}A along with the low-$T$ heat capacity.
These data are consistent with a second-order mean-field transition with the critical exponent $\beta=0.48(4)$.
Thus, the magnetic structure of CeAlSi belongs to a single irreducible representation (irrep) of the $I4_1md$ space group.
The combination of our symmetry analysis (see Sec. M3 of Supplementary Material) with the observation of several (00L) peaks allows us to conclude that CeAlSi orders in the $\Gamma_5$ manifold, where all spins lie in the $ab$-plane (Fig.~\ref{fig:NM}C).

As illustrated in the inset of Fig.~\ref{fig:NM}C, the $\Gamma_5$ manifold allows for a complete decoupling of the Ce spins between the adjacent (0,0,$z$+1/4) layers.
We define $\mathbf{S_1}$ to be the Ce spin at (0,0,0) and $\mathbf{S_2}$ to be that at (0,1/2,1/4).
Intensity of the neutron Bragg peaks with $\mathbf{k}$~=~(000) and (110) ordering vectors is proportional to $\mathbf{S}_{1}$+$\mathbf{S}_{2}$ and $\mathbf{S_{1}}$-$\mathbf{S_{2}}$, respectively.
Thus, the observation of both the ordering vectors in Figs.~\ref{fig:NM}C,D suggests that both $\mathbf{S}_{1}$+$\mathbf{S}_{2}$ and $\mathbf{S_{1}}$-$\mathbf{S_{2}}$ are finite, so that the angle between $\mathbf{S}_{1}$ and $\mathbf{S}_{2}$, defined by $\theta=\cos^{-1}$($\frac{\mathbf{S_1}\cdot\mathbf{S_2}}{\Vert\mathbf{S_2}\Vert \Vert\mathbf{S_2}\Vert}$), must be nonzero.
Detailed refinement of the spin structure was then performed against 40 symmetrically distinct Bragg peaks collected at both 1.4~K and 10~K in zero-field, see Sec. M3 in Supplementary Material for details.
Assuming $\Vert\mathbf{S_1}\Vert=\Vert\mathbf{S_2}\Vert$, the refinement suggests a moment size of 1.2(2)$\mu_B$ and $\theta=70(30)^\circ$,  confirming the in-plane noncollinear FM order in CeAlSi (inset of Fig.~\ref{fig:NM}D).

We performed scanning SQUID microscopy~\cite{sochnikov_nonsinusoidal_2015,sochnikov_direct_2013,gardner_scanning_2001} to visualize the FM domain structure of CeAlSi.
The images in Figs.~\ref{fig:NM}E,F were obtained by scanning a SQUID sensor over the $ab$-surface of a polished crystal to measure the out-of-plane stray field from the in-plane domains.
Although the domains are small under zero-field-cooling (Fig.~\ref{fig:NM}E), a weak in-plane field of a few Gauss is enough to generate large in-plane FM domains that are hundreds of microns across (Fig.~\ref{fig:NM}F).
The development of large domains is also implied by the selection of a single domain revealed by neutron diffraction. Under a small in-plane field $\mathbf{H}\|$~[110], the system selects a single domain with $\mathbf{M}\|$~[110] among all symmetrically equivalent directions. As a result, the vector $\mathbf{S_{1}}$-$\mathbf{S_{2}}$ only points along [1-10] and the $\mathbf{Q}$~=~(1-10) Bragg peak is suppressed accordingly, as seen in Fig.~\ref{fig:UC}D when a field of 0.1~T is applied in the [110] direction.
Magnitude of the observed DC flux is on the order of a few $\Phi_0$, consistent with the remnant magnetization determined from the $c$-axis bulk magnetization measurements.
According to our estimates based on the remanent $a$-axis magnetization, if the domains were to have flipped magnetization from the in-plane to the out-of-plane direction, it would have produced DC signal on the order of hundreds of $\Phi_0$, which is clearly not the case in Figs.~\ref{fig:NM}E,F, see Sec. M4 of Supplementary Material for details).
The picture that emerges from our neutron scattering and scanning SQUID measurements in CeAlSi is that of a noncollinear in-plane FM order with large domains.

\section{\label{sec:theory}Band Structure, Shifted Weyl Nodes, and Intrinsic Anomalous Hall Conductivity}
\begin{figure*}
\centering
\includegraphics[width=\textwidth]{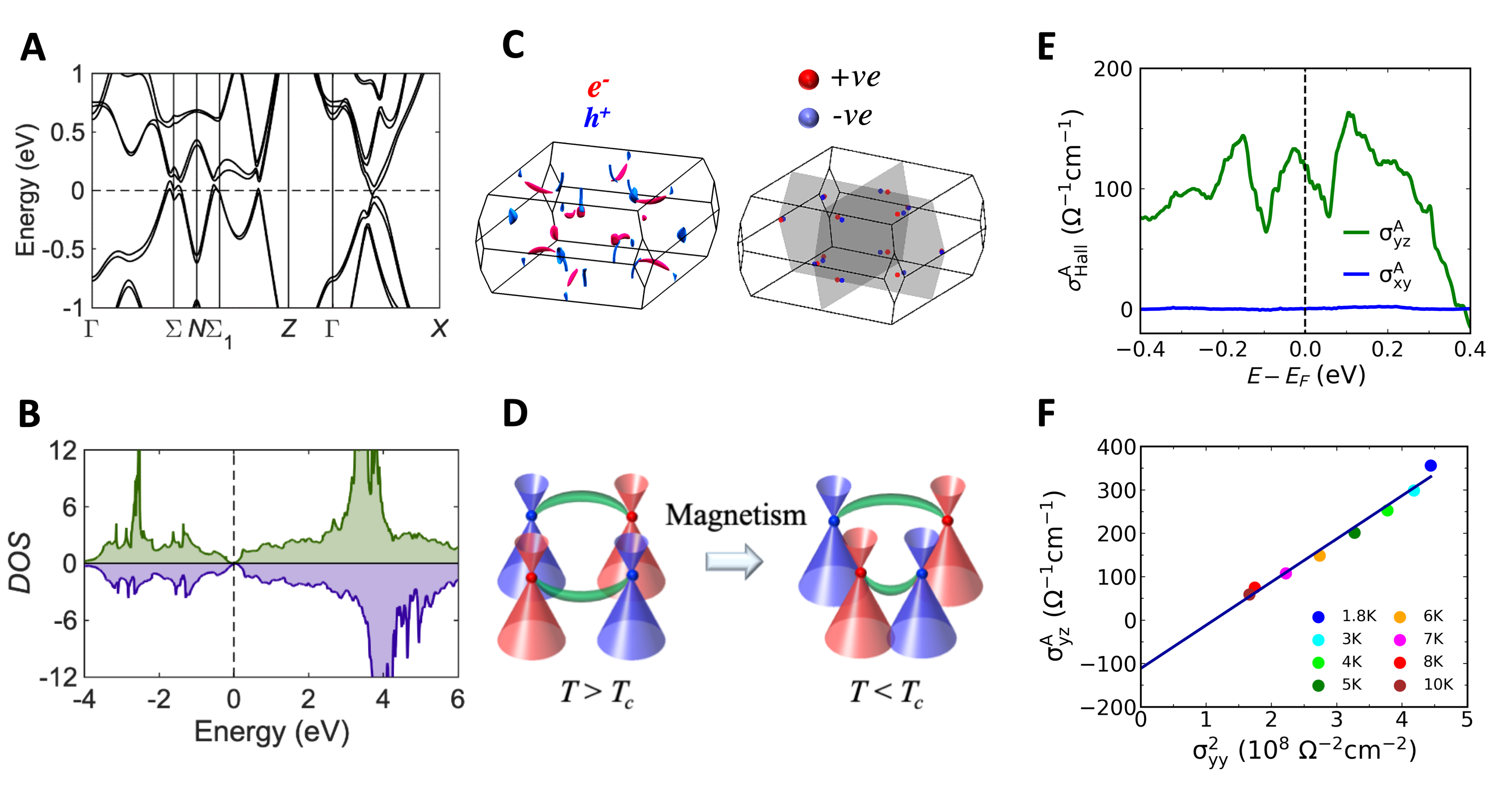}
\caption{\label{fig:DFT}
\textbf{Band structure and Weyl nodes.}
(\textbf{A}) Band structure of CeAlSi including spin-orbit coupling.
(\textbf{B}) Partial density of states for the majority (green) and minority (purple) spin channels.
(\textbf{C}) Electron (red) and hole (blue) pockets (left) and the Weyl nodes (right) are shown in the BZ.
(\textbf{D}) Effects of breaking the inversion (left) and time-reversal (right) symmetries on the positions of Weyl nodes are shown schematically.
(\textbf{E}) Theoretical values of the anomalous Hall conductivity with the field oriented in-plane ($\sigma_{yz}^A$) and out-of-the-plane ($\sigma_{xy}^A$).
(\textbf{F}) Scaling behavior $\sigma_{yz}^A\propto \sigma_{xx}^2$ in CeAlSi, confirming an intrinsic AHE.
}
\end{figure*}
Band structure of CeAlSi (Fig.~\ref{fig:DFT}A) consists of small hole and electron pockets with a nearly vanishing density of states (DOS) at $E_F$ (Fig.~\ref{fig:DFT}B).
The DOS in the majority and minority spin channels peaks at different energies (Fig.~\ref{fig:DFT}B) and leads to FM ordering.
The residual electron and hole pockets are illustrated in Fig.~\ref{fig:DFT}C, which also shows the 12 pairs of Weyl nodes next to the $k_x=0$ and $k_y=0$ mirror-planes.
We denote the 4 pairs of nodes located on the $k_z=0$ plane as $W_1$, and the other 8 as $W_2$.
The $W_1$ nodes are $80-120$~meV away from $E_F$ but the $W_2$ nodes lie within $25$~meV of the $E_F$, see Sec. M5 in Supplementary Material for details.
All $W_1$ and $W_2$ Weyl fermions exhibit linear energy dispersions in all $k$-directions, suggesting that CeAlSi is a type-I WSM, see Sec. M5 and Fig. M5 of Supplementary Material for details.
This is different from the case of the related material CeAlGe that hosts both type-I and type-II Weyl nodes~\cite{chang_magnetic_2018,xu_discovery_2017} driven by the stronger spin-orbit coupling of Ge and the slightly different Wyckoff site coordinates.
Note that the Weyl nodes in CeAlSi result from a broken inversion symmetry ($\mathcal{I}$) and the effect of breaking the time-reversal symmetry ($\mathcal{T}$) at $T<T_C$ is to shift the positions of the Weyl nodes in the BZ (Fig.~\ref{fig:DFT}D)~\cite{yang_quantum_2011,chang_magnetic_2018}.
CeAlSi is thus a new FM-WSM, in sharp contrast to the centrosymmetric systems such as Co$_3$Sn$_2$S$_2$~\cite{liu_giant_2018,wang_large_2018}, Fe$_3$GeTe$_2$~\cite{kim_large_2018}, and the Heusler alloys~\cite{wang_time-reversal-breaking_2016} where the Weyl nodes result from the broken $\mathcal{T}$.

We calculated the anomalous Hall conductivity (AHC) along the easy ($\sigma_{yz}^A$) and hard ($\sigma_{xy}^A$) axes as a function of the Fermi energy in Fig.~\ref{fig:DFT}E~\cite{yao_first_2004}.
Magnitude of the theoretical AHC along the easy axis in Fig.~\ref{fig:DFT}E is comparable to the corresponding experimental values ($\sigma_{xy}^A = -\rho_{xy}^A/\rho_{xx}^2$ ; $\rho_{xy}^A=\rho_{xy}-R_0H$)~\cite{tian_proper_2009,yang_transition_2020} in Fig.~\ref{fig:DFT}F.
The scaling behavior between $\sigma_{xy}^A$ and $\sigma_{xx}^2$ (Fig~\ref{fig:DFT}F) indicates the presence of intrinsic and extrinsic contributions to the AHE~\cite{nagaosa_anomalous_2010}, where the y-intercept codes the intrinsic contribution and the scaling with $\sigma^2_{xx}$ represents the extrinsic contribution\cite{tian_proper_2009,yang_transition_2020}.
Note that, according to the DFT, we do not expect an AHC ($\sigma_{xy}^A$~=~0) along the magnetic hard-axis $H\|c$.
Therefore, the observation of a loop-shaped Hall signal with $H\|c$ is a novel electromagnetic response as discussed further below.

\section{\label{sec:hall}Anisotropic Anomalous Hall Effect}
\begin{figure*}
\centering
\includegraphics[width=\textwidth]{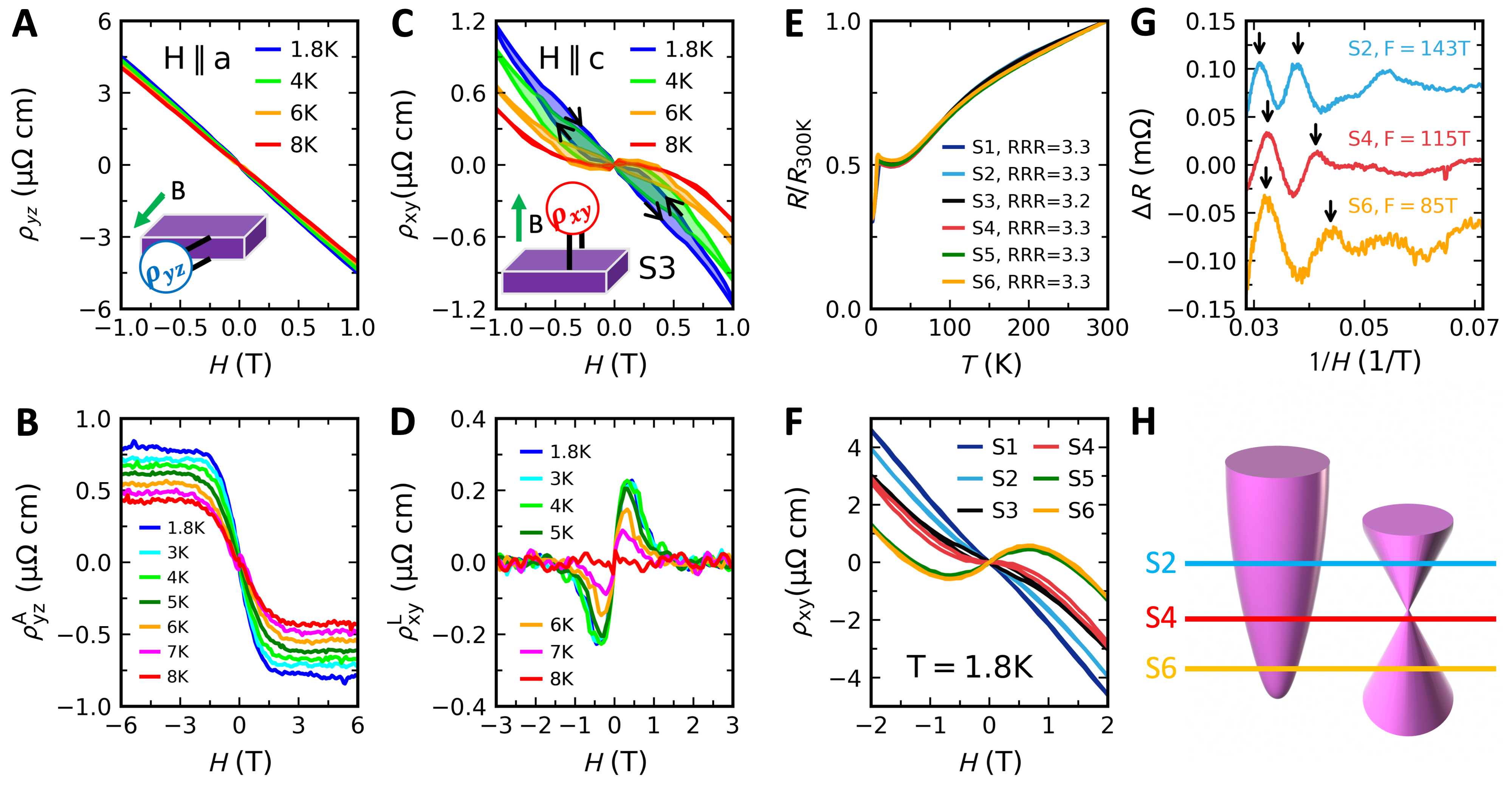}
\caption{\label{fig:AHE}
\textbf{Magnetic anisotropy and the loop Hall effect.}
(\textbf{A}) Hall resistivity $\rho_{yz}$ measured at several temperatures with the field along the magnetic easy axis ($H\|a$).
(\textbf{B}) Anomalous Hall resistivity $\rho_{yz}^A$ derived from $\rho_{yz}$.
(\textbf{C}) Hall resistivity $\rho_{xy}$ measured at several temperatures with the field along the hard axis ($H\|c$).
(\textbf{D}) Loop Hall resistivity $\rho_{xy}^L$ derived from $\rho_{xy}$.
(\textbf{E}) Normalized resistivity plotted as a function of temperature in samples S1-S6 with similar residual resistivity ratios.
(\textbf{F}) Hall resistivity $\rho_{xy}$ of samples S1-S6 measured at $T=1.8$ K.
(\textbf{G}) Quantum oscillations (QOs) in samples S2, S4, and S6, showing the evolution of the oscillation period.
(\textbf{H}) Schematic band structure of CeAlSi illustrating an electron band, a Weyl crossing, and the Fermi levels of samples S2, S4 and S6. The decreasing QO frequency seen in panel (\textbf{G}) originates from the electron pocket (left) as the Fermi level shifts in samples S2, S4 and S6.
}
\end{figure*}
CeAlSi displays two different anomalous Hall responses when measured under in-plane (easy-axis) and out-of-plane (hard-axis) magnetic fields; such a concurrence is quite unusual and has been reported only in few cases \cite{takahashi2018anomalous}.
We start by examining the Hall resistivity $\rho_{yz}$ as a function of the in-plane field $H\|a$ (Fig.~\ref{fig:AHE}A).
We separate the conventional and anomalous Hall signals by fitting the data at $H>1$~T to $\rho_{yz}=R_0H+\rho^A_{yz}$.
The conventional Hall effect ($R_0H$) has a slope $R_0=-3.9$~$\mu\Omega$cmT$^{-1}$ corresponding to a small electron concentration $n_e=-1.6\times 10^{20}$~cm$^{-3}$ (0.0003 electrons per unit cell), consistent with the small value of DOS at the $E_F$ in Fig.~\ref{fig:DFT}B.
The anomalous Hall resistivity $\rho^A_{yz}$ is plotted as a function of field in Fig.~\ref{fig:AHE}B.
Note that $\rho^A_{yz}$ does not exactly follow the magnetization (Fig. 1D) like other conventional AHE materials, which is expected in systems with noncollinear spin texture such as Pr$_2$Ir$_2$O$_7$\cite{machida2007unconventional,balicas2011anisotropic}.
The anomalous Hall conductivity calculated from $\sigma_{yz}^A=-\rho_{yz}^A/\rho_{xx}^2$ is plotted in Figs.~\ref{fig:UC}E and ~\ref{fig:DFT}G.
Magnitude of $\sigma_{yz}^A$ is in agreement with the DFT results in Fig.~\ref{fig:DFT}E.

Next, we discuss the Hall resistivity $\rho_{xy}$ as a function of the out-of-plane field $H\|c$ (hard-axis) in Fig.~\ref{fig:AHE}C, where an unusual loop is observed.
This loop corresponds to different traces of $\rho_{xy}(H)$ between the field sweeps in the positive and negative directions (arrows in Fig.~\ref{fig:AHE}C).
It extends over a region of $\pm 2$~T, two orders of magnitude larger than the magnetic coercive field (70~Oe, inset of Fig.~\ref{fig:UC}D).
Notice that the loop-shaped Hall effect (LHE) does not scale with magnetization $M(H)$ and appears only when measured along the magnetic hard axis, unlike the AHE that follows the magnetization curve ($\rho^A_{yz}=R_SM(H)$) and appears when the field is parallel to the easy axis.
In order to study the temperature dependence of the LHE, we subtract the positive field sweep from the negative sweep and plot the loop Hall resistivity as $\rho_{xy}^L= \rho_{xy}(3\to  -3~\textrm{T})-\rho_{xy}(-3\to 3~\textrm{T})$ at several temperatures (Fig.~\ref{fig:AHE}D).
The loop Hall conductivity $\sigma^L_{xy}$ in Figs.~\ref{fig:UC}E,F was calculated as $\sigma^L_{xy}=\rho^L_{xy}/\rho^2_{xx}$, see Sec. M6 in Supplementary Material for details of the $\rho_{xx}$ data.

In order to explore the link between the LHE and Weyl nodes, we measured samples with different separations between the Fermi level and the Weyl nodes.
For this purpose, we selected six samples (S1-S6) with comparable residual resistivity ratios $RRR=R(300\textrm{K})/R(2\textrm{K})$ (Fig. \ref{fig:AHE}E).
Slight off-stoichiometry of Si and Al in our samples (Sec. M7 of Supplementary Material) causes a shift of the Fermi level relative to the Weyl nodes~\cite{guo_evidence_2018}. 
Variations in the $E_F$ between the samples is evident in Fig.~\ref{fig:AHE}F, which shows three categories of Hall curves:
a linear $\rho_{xy}$ with negative slope in samples S1 and S2; a moderately nonlinear $\rho_{xy}$ with negative slope at all fields in S3 and S4; and, a strongly nonlinear $\rho_{xy}$ with positive slope at low fields and negative slope at high fields in S5 and S6.
%
%
Since the slope of $\rho_{xy}$ is related to the sign of charge carriers, 
we adduce that $E_F$ crosses only electron pockets in samples S1 and S2, nearly crosses another hole pocket in S3 and S4, and crosses both the electron and hole pockets in S5 and S6 as illustrated in Fig.~\ref{fig:AHE}H.
The LHE is observed only in S3 and S4 where the $E_F$ lies near the crossing of electron and hole bands, i.e. near the Weyl node (Fig.~\ref{fig:AHE}H).

In order to confirm the scenario of Fig.~\ref{fig:AHE}H, we used Schubnikov-de Haas (SdH) oscillations to locate the $E_F$ with respect to the Weyl nodes along the lines of prior work on Weyl and magnetic semimetals~\cite{schonemann_fermi_2017,yang_interplay_2018}.
Figure~\ref{fig:AHE}G shows quantum oscillations for magnetic field between 15 and 33~T in samples S2, S4, and S6.
The frequency of SdH oscillations, $F=A\left(\frac{\hbar}{2\pi e}\right)$, is proportional to the extremal orbit area $A$, and it will change as we shift the $E_F$ in the band structure. The $E_F$ for each sample can then be pinned down by matching experimental and theoretical frequencies of the electron pocket (the left portion in Fig.~\ref{fig:AHE}H).
Through such an analysis, we obtain $E_F$ values for samples S2, S4, and S6 to lie 32, 23, and 12 meV above the DFT-calculated value, respectively; see Sec. M8 of Supplementary Material for details.
When we compare these $E_F$ values to the energies of Weyl nodes, we find that all Weyl nodes lie away from the $E_F$ in samples S2 and S6, but a set of $W_2^2$ Weyl nodes is located within $1$~meV of the $E_F$ in sample S4, as illustrated in Fig.~\ref{fig:AHE}H; see also Tables M2 and M4 of Supplementary Material.
Thus, we conclude that the LHE is observed only in samples where the  $E_F$ nearly crosses the Weyl nodes.

\section{\label{sec:outlook}Outlook}
In summary, CeAlSi is a unique noncentrosymmetric FM-WSM with an in-plane noncollinear FM order and novel anisotropic anomalous Hall responses along the easy and hard magnetic axes.
In particular, CeAlSi exhibits the LHE which appears when the applied field lies along the hard axis. The LHE does not scale with either the field or the magnetization and is deeply connected with the Weyl nodes.
The LHE is distinct from the THE~\cite{PhysRevLett.102.186602,PhysRevLett.106.156603,matsuno_interface-driven_2016} because the magnetic structure of CeAlSi may not support spin chirality or a skyrmion phase.
In order to gain insight into the LHE in CeAlSi, we consider Nd$_2$Ir$_2$O$_7$, which also exhibits loop-shaped signals in magnetoresistance and Hall resistivity \cite{ueda_anomalous_2014,disseler_magnetization_2013}. 
Nd$_2$Ir$_2$O$_7$ hosts an all-in-all-out magnetic order of null spin chirality and requires an explanation other than the THE for its loop responses.
Recently, it was proposed that Nd$_2$Ir$_2$O$_7$, despite having an insulating ground state, is very close to a WSM phase and that slight doping or external pressure will turn it into a WSM \cite{witczak2012topological,ueda2012variation,ueda2018spontaneous}.
%
As a result, topological Fermi arcs in Nd$_2$Ir$_2$O$_7$ projected from the Weyl nodes on the magnetic domain walls interact to form exotic surface states (SSs); these topological Fermi-arc-induced (FAI) SSs survive the annihilation of Weyl nodes in the insulating regime \cite{yamaji_metallic_2014}.
%
The FAI SSs have been mapped out in Nd$_2$Ir$_2$O$_7$ by impedance spectroscopy \cite{ma2015mobile}, and can serve as special conducting channels responsible for the anomalous loop responses \cite{disseler_magnetization_2013,ueda_anomalous_2014,
yamaji_metallic_2014}.

Keeping the preceding discussion of the FAI SSs in mind, we compare and contrast Nd$_2$Ir$_2$O$_7$ and CeAlSi to gain insight into the origin of the LHE in CeAlSi as follows.
\begin{enumerate}
\item Nd$_2$Ir$_2$O$_7$ is an overall AFM system with an all-in-all-out magnetic order, whereas CeAlSi hosts a non-collinear FM order. FAI SSs, however, only require the presence of the magnetic domain walls and the proximity of a WSM phase, and can thus be expected also in CeAlSi.
\item Although Nd$_2$Ir$_2$O$_7$ is insulating whereas CeAlSi is semimetallic, FAI SSs can exist in both materials. In Nd$_2$Ir$_2$O$_7$, FAI SSs are remnant traces of the Fermi arcs in the system before it becomes insulating, while in CeAlSi, they are the Fermi arcs connecting the bulk Weyl nodes.
\item The loop response in Nd$_2$Ir$_2$O$_7$ appears in both magnetoresistance ($\rho_{xx}$) and Hall resistivity $\rho_{xy}$, whereas in CeAlSi it only appears in $\rho_{xy}$. Generally, $\rho_{xx} \sim \sum_i^n \frac{\sigma_i}{1+\mu_i^2 B^2}$, where the summation extends over all conducting bands. Since $\sigma_i = n_ie_i\mu_i$ is always positive, $\rho_{xx}$ is dominated by the bands with large carrier densities $n$.  Since Nd$_2$Ir$_2$O$_7$ is insulating, the FAI SSs provide the only conducting channels and dominate $\rho_{xx}$ and lead to the loop-shaped behavior. CeAlSi, in contrast, is metallic and its topological SSs fail to show a loop response in $\rho_{xx}$ because the small density of states associated with these SSs is overwhelmed by the contribution from the bulk bands. On the other hand, note that $\rho_{xy} \sim \sum_i^n \frac{\sigma_i \mu_i}{1+\mu_i^2 B^2}$ and it can, therefore, be either positive or negative depending on the sign of the carriers. In Nd$_2$Ir$_2$O$_7$, the FAI SSs being the only carriers, they also drive $\rho_{xy}$ and yield a loop response. In CeAlSi, the electron and hole (bulk) contributions to $\rho_{xy}$ nearly cancel (Fig. \ref{fig:DFT}C) and, as a result, the topological SSs control the behavior of $\rho_{xy}$ and drive its loop response in CeAlSi. This argument is consistent with our quantum oscillation results, which reveal an enhanced SS contribution (LHE) in the CeAlSi samples in which the Fermi energy lies close to the Weyl nodes.
\end{enumerate}

CeAlSi will not only be amenable to ARPES studies due to its metallicity but it will also be suitable for device engineering and tuning of the Fermi arcs~\cite{ilan_pseudo-electromagnetic_2020}. CeAlSi would thus provide an interesting materials platform for exploring the physics of Weyl nodes and how these nodes are connected with the exotic electromagnetic responses of topological materials.

\begin{acknowledgments}
We thank Chunli~Huang, Hiroaki~Ishizuka, Bohm-Jung~Yang and Ying~Ran for helpful discussions.
F.T. acknowledges funding by the National Science Foundation under Award No. NSF/DMR-1708929.
The work at Northeastern University was supported by the US Department of Energy (DOE), Office of Science, Basic Energy Sciences grant number DE-SC0019275 and benefited from Northeastern University's Advanced Scientific Computation Center (ASCC) and the NERSC supercomputing center through DOE grant number DE-AC02-05CH11231.
Neutron scattering was supported as part of the Institute for Quantum Matter, an Energy Frontier Research Center funded by the U.S. Department of Energy, Office of Science, Basic Energy Sciences under Award No. DE-SC0019331.
J.G. and C.B. were supported by the Gordon and Betty Moore foundation under the EPIQS program GBMF9456.
 A portion of this research used resources at the High Flux Isotope Reactor and Spallation Neutron Source, a DOE Office of Science User Facility operated by the Oak Ridge National Laboratory.
 The National High Magnetic Field Laboratory is supported by National Science Foundation through NSF/DMR-1644779 and the State of Florida.
 The work by I.S. was in part supported by the US Department of Defense, and the US State of Connecticut.
 B.X. and J.F. were supported through graduate assistantship provided by the University of Connecticut's College for Liberal Arts and Sciences.
 We acknowledge the support of the National Institute of Standards and Technology, U.S. Department of Commerce.
 Certain commercial equipment, instruments, or materials (or suppliers, or software, etc.) are identified to foster understanding.
 Such identification does not imply recommendation or endorsement by the National Institute of Standards and Technology, nor does it imply that the   materials or equipment identified are necessarily the best available for the purpose.

The authors declare no competing financial interests.

H.-Y. Y. and B. S. contributed equally to this work. H.-Y. Y. grew the crystals, performed magnetization and transport experiments. H.-Y. Y. and D. E. G. performed high-field experiments.
 B. S. performed first-principles calculations and theoretical analysis with assistance and guidance from C.-Y. H., W.-C .C.,  S.-M. H., B. W., H. L., and A. B..
 J. G. and C. L. B. performed neutron scattering, B. L. and D. H. T. performed SHG experiments, F. B. analyzed X-ray data.
 I. S., B. X., and J. F. performed the scanning SQUID microcopy.
 F. T. and A. B. conceived the research. All authors discussed the results and contributed to writing the manuscript.
\end{acknowledgments}

\appendix*
\section{Methods}
\subsection*{Crystal Growth}
CeAlSi single crystals were grown by a self-flux method in both regular alumina crucibles and the Canfield crucible sets~\cite{canfield_use_2016}.
Both methods produced a similar crystal quality based on the PXRD, SHG, EDX and resistivity measurements.
In both methods, the starting materials were weighed in the ratio Ce:Al:Si = 1:10:1, placed inside a crucible in an evacuated quartz tube, heated to 1000~$^\circ$C at 3~$^\circ$C/min, stayed at 1000~$^\circ$C for 12~h, cooled to 700~$^\circ$C at 0.1~$^\circ$C/min, stayed at 700~$^\circ$C for 12~h, and centrifuged to decant the residual Al flux.

\subsection*{Band Structure}
Density functional theory (DFT) calculations were performed using the experimental lattice parameters ($a = 4.252$~\AA; $c = 14.5801$~\AA) and the projector-augmented-wave (PAW) method implemented in the Vienna ab-initio simulation package (VASP)~\cite{kresse_efficient_1996}.
The exchange-correlation effects were included using the generalized gradient approximation (GGA). The spin-orbit coupling (SOC) was included self-consistently~\cite{kresse_ultrasoft_1999,perdew_generalized_1996}.
An on-site Coulomb interaction was added for Ce $f$-electrons within the GGA+U scheme with $U_{\textrm{eff}} = 6$~eV.
A Wannier tight-binding Hamiltonian was obtained from the ab-initio results using the VASP2WANNIER90 interface, which was subsequently used in our topological properties calculations~\cite{marzari_maximally_1997}.

\subsection*{Transport, Heat capacity, and Magnetization Measurements}
Electrical resistivity was measured with the standard four-probe technique and the heat capacity was measured with the relaxation time method in a Quantum Design Physical Property Measurement System (PPMS) Dynacool. Magnetic heat capacity $C_m$ was obtained by first measuring the heat capacity of non-magnetic LaAlSi, and then subtracting it from the heat capacity of CeAlSi.
DC magnetization experiments were conducted on the vibrating sample magnetometer in a Quantum Design MPMS3.
The high-field experiments were performed using a 35~T DC Bitter magnet and a $^3$He fridge with base temperature of 300~mK at the MagLab in Tallahassee.
Comparison of the quantum oscillation frequencies between theory and experiment was carried out by using the DFT-generated bxsf file and the program SKEAF \cite{julian2012numerical}.

\subsection*{Neutron Diffraction}
The nuclear structure of CeAlSi was characterized by a single-crystal time-of-flight experiment at 100~K on TOPAZ at the Oak Ridge National Lab.
A 3D diffraction map was acquired from 14 different sample positions allowing measurements of 6946 Bragg peaks where the nuclear structure factors were extracted following the method of Schultz \textit{et al.}~\cite{schultz_integration_2014}.
Structural refinements were performed using GSAS-II~\cite{toby_gsas-ii_2013}.
The magnetic structure was determined by diffraction experiments at the NIST Center for Neutron Research.
The magnetic structure factors were determined using the thermal triple-axis spectrometer BT-7 by collecting rocking scans at various Bragg positions with incident and scattered neutron energies of 14.7~meV.
Two single crystals were inserted in a top-loading CCR and a 7~T magnet to measure Bragg peaks in both the (H0L) and (HHL) planes.
The order parameter measurement in Fig.~\ref{fig:UC}F was performed with the SPINS spectrometer using 3.7~meV incident and scattered neutrons.

\subsection*{Second Harmonic Generation}
The SHG data in Fig.~\ref{fig:UC}B were taken at normal incidence on the [101] face of as-grown crystals for incoming(outgoing) wavelength of 1500(750)~nm as a function of the incoming field polarization and measured for emitted light polarized parallel to the [010] crystalline axis~\cite{lu2019fast}.
In this geometry, all bulk contributions to the SHG signal from a $I4_1/amd$ space group are forbidden.

\subsection*{Scanning SQUID Imaging}
We used scanning SQUID susceptometers with two gradiometric field coils and pickup loops~\cite{sochnikov_nonsinusoidal_2015}.
The SQUID pickup loop and the field-coil average radii were 3.25 and 7~$\mu$m formed from Nb lines of 0.5 and 1~$\mu$m width, respectively.
The scanning SQUID apparatus was housed in a closed-cycle Montana Instruments Fusion cryostat (Bozeman, Montana, USA) with the cryostat base temperature of 3 K.

\bibliography{Yang_19Dec2020}

\end{document}